# Queuing Methodology Based Power Efficient Routing Protocol for Reliable Data Communications in Manets

## Giddaluru Madhavi[a] and M.K. Kaushik[b,*]

[a]*Information Science And Engineering, The Oxford College Of Engineering , Bengaluru, India*
[b]*Department Of Electrical Engineering, BITS-Pilani, Hyderabad Campus, Hyderabad, India*

A mobile ad hoc network (MANET) is a wireless network that uses multi-hop peer-to- peer routing instead of static network infrastructure to provide network connectivity. MANETs have applications in rapidly deployed and dynamic military and civilian systems. The network topology in a MANET usually changes with time. Therefore, there are new challenges for routing protocols in MANETs since traditional routing protocols may not be suitable for MANETs. In recent years, a variety of new routing protocols targeted specifically at this environment have been developed, but little performance information on each protocol and no realistic performance comparison between them is available. This paper presents the results of a detailed packet-level simulation comparing three multi-hop wireless ad hoc network routing protocols that cover a range of design choices: DSR, NFPQR, and clustered NFPQR. By applying queuing methodology to the introduced routing protocol the reliability and throughput of the network is increased.

## 1. Introduction

Today's Internet has been developed for more than forty years. Recently many network researchers are studying networks based on new communication techniques, especially wireless communications. Wireless networks Allow hosts to roam without the constraints of wired connections. People can deploy a wireless network easily and quickly. End users can move around while staying connected to the network. Wireless networks play an important role in both military and civilian systems. Handheld personal computer connectivity, notebook computer connectivity, vehicle and ship networks, and rapidly deployed emergency networks are all applications of this kind of network.

Mobile Ad hoc Network(MANET) is a collection of wireless mobile nodes with dynamic changing topology forming a temporary network without infrastructure or centralized administration [1]. Mobile Ad hoc Network has become an active research area in the domain of wireless networking because of their distinctive advantages which includes easier set up, saving in hardware cost [2 ]. Each node can move independently in any direction and also act as a router for communication between nodes which are not within radio distance. The Mobile Ad hoc Network, because of its fast and economically less demanding service, find applications military, collaborative and distributed computing, emergency operations, wireless mesh networks, wireless sensor networks, hybrid wireless network architectures and educational environments .In MANET privacy of the nodes are ensured by Anonymous communication which also enhances the security of the network[3].

Traditional routing protocols used in wired networks ar e ineffective for ad hoc networks because of the intrinsic qualities of wireless media and the dynamic changing topology of the network [4] . Most of the proposed routing protocols in an Ad hoc network are enhancement of their wired counterpart and can be broadly classified into Proactive, Reactive and Hybrid routing protocol [5,12]. Proactive routing protocols are also known as table driven routing protocols. These protocols create routing table as the network is formed and dynamically updates the routing table when the network topology changes. Examples of proactive protocols are Destination-Sequenced Distance- Vector (DSDV) [6, 13] Optimized link state routing protocol (OLSR ) [7] and Cluster head Gateway Switch Routing Protocol (CGSR) [8]. When the size of the network is very large,
the size of the corresponding routing table is also large which is a disadvantage for memory constraint nodes. However these problems have been overcome in some of the proactive routing protocols including OLSR and CGSR. Reactive routing protocols are also known a s on demand routing protocols. Reactive protocols discover route only when data is to be transmitted between two nodes. Examples of re active protocols are Dynamic Source Routing (DSR) [9,11], Ad hoc On demand Distance Vector (AODV) [10 , 12]. Typical problems in reactive routing protocols include higher latency for route discovery and network congestion due to excessive flooding. Hybrid routing protocols combine the advantage of both proactive and reactive routing. Initially routes are established using some proactive technique and subsequently updated as and when required using reactive technique. The choice for one or the other method needs predetermination for typical cases. Some of the issues
in hybrid routing protocols are high latency for new route discovery.

Hosts and routers in a wireless network can move around. Therefore, the network topology can be dynamic and unpredictable. Traditional routing protocols used for wired networks cannot be directly applied to most wireless networks because some common assumptions are not valid in this kind of dynamic network. For example, one assumption is that a node can receive any broadcast message sent by others in the same subnet. However, this may not be true for nodes in a wireless mobile network. The bandwidth in this kind of network is usually limited. Thus, this network mode l introduces great challenges for routing protocols. Many mobile ad hoc network (MANET) routing protocols have been proposed. Past work focused on designing new protocols, comparing existing protocols, or improving protocols before standard MANET routing protocols are defined. Most research in this field is based on simulation studies of the routing protocols of interest in arbitrary networks with certain traffic profiles. However, the simulation results from different research groups are not consistent. This is because of the lack of consistency in MANET routing protocol models and application environments including networking and user traffic profiles. Therefore, simulation scenarios used in past studies are not fair for all protocols and their conclusion s cannot be generalized. Furthermore, it is difficult for one to choose a proper routing protocol for a given MANET application.

## 2 Literature Review

Recently, the increased demand for ubiquitous internet connectivity and broadband internet service has spurred the need for new innovative wireless technologies [2]. WMNs are one such upcoming technology that offer wireless broadband internet connectivity that would provide varied functionalities. They offer cost-effective and flexible solution for extending broadband services to the residential areas without any necessity for line-of-sight communication. WMNs are formed by a set of mesh routers (MRs), among which a small subset is directly connected to the wired network called the Internet Gateway (IGW). WMNs are based on the ad hoc networking paradigm [1] and thus adopt a self-configurable and self-healing approach. For a network to have a reliable communication routing protocol is an important requirement.

Routing Protocols in Ad-Hoc networks have been classified in many ways[3], based on routing strategy and network structure. According to routing strategy, the routing protocols can be categorized as table-driven and source initiated, while depending on the network structure these are classified as flat routing, hierarchial routing and geographic position assisted routing. It should be mentioned at this point that owing to the dynamic link topology, there are both unipath and multipath routing protocols. Most multipath routing protocols are built upon unipath routing protocols, hence we restrict our discussion to mostly unipath routing protocols.

A lot of the routing protocols mainly make use of the flooding technique for route discovery which we briefly mention as follows: A sender S broadcasts data packet P to all its neighbors, and each node receiving P forwards P to its neighbors - In this way, packet P reaches the destination D provided a path from S to D exists. Node D does not forward the packet. We describe two routing protocols that make use of this technique, namely Dynamic Source Routing (DSR) [4] and Ad-hoc On-Demand Distance Vector Routing (AODV) [5]. DSR uses source routing instead of relying on the routing table at each intermediate device. DSR uses a combination of Route Request (RREQ), Route Reply (RREP) and Route Error (RERR) messages to establish a route from source to destination. In addition DSR also uses route caching to speed up route discovery. The main advantage of DSR is that a route to a destination is established only when it is required to, and a single route discovery often discovers routes to other destination s in the process. However, the main drawbacks of DSR are increased size of packet header with the length of the route, stale caches, etc. AODV is a reactive(on-demand) protocol which borrows the advantageous concepts from DSR and DSDV (Destination Sequenced Distance Vector Routing) (We mention this later) like on-demand route discovery and route maintenance, and the usage of node sequence numbers from DDSV. The actual working of AODV is beyond our scope. See [5] for a good description. AODV is one of the most successful routing protocols in MANETs mainly because of its desirable features like Minimal Space Complexity and Maximum Bandwidth Utilization.

We now look at another widely deployed protocol called Destination Sequenced Distance Vector Routing protocol (DSDV) [6] . DSDV is a proactive (means routes are independent of traffic pattern) protocol which is a modification of the conventional Bellman-Ford algorithm. The important feature of DSD V is the use of sequence numbers for the routing table entries. These sequence numbers are generated by the destination stations and routing tables at each node are synchronized by each node advertising its routing table information to its neighbors frequently. Forwarding decisions are made based on the sequence numbers. The main advantages of DSDV are that it guarantees loop-free paths and always maintains the best path to destination rather than multiple paths. However, routing table up dates are costly, and there is no support for multi-path routing.

More recently, because of the rapid advances in Global Positioning Systems(GPS), Geographic Routing Protocols are becoming popular. One such protocol is the Geographic Distance Routing (GEDIR) [7].

In GEDIR, location of the destination node is assumed to be known, and each node knows the locations of its neighbors and forwards a packet to the neighbor closest to the destination. Many interesting advances have been made in Geographic Routing like Routing based on Virtual Co-ordinates [9], Routing without Location Information [8], etc.

In addition to Reactive and Proactive routing protocols, we also have Hybrid schemes like the Zone Routing Protocol (ZRP) [10] which proactively maintain state information for links within a short distance, say d, from any given node (Intra-Zone Routing) and uses a route discovery protocol for determining routes to a nodes at a distance greater than d (Inter-Zone Routing).

Most protocols we have seen so far use some kind of flooding. There are few protocols like the Link Reversal Algorithm and Temporally-Ordered Routing Algorithm (TORA) which try to avoid/reduce flooding behaviour. See [11] and [12] respectively for discussions. Routing Protocols which define the optimization criteria as a function of the energy consumption are called Power-Aware Routing Protocols [13]. The idea is to assign each link a weight which is a function of energy consumed when transmitting a packet on that link. The goal is to route through paths with minimum weight. Such protocols are usually built on top of existing protocols like DSR.

The present work gives idea about the congestion control avoidance in mobile adhoc networks. Once the Routing protocol is said to be reliable then other issues also should be kept in mind like node behavior and timing delays also. In order to come up with good communication system with addition to these routing protocols one mathematical analogy called queuing model is also applied then we can visualize all the parameters properly.

Due to the dynamicity of the link factors there are various issues been observed, in which congestion is one of the problem. Congestion is caused when the offered load to the network is more than the available resources. To overcome the congestion problem in mobile adhoc network a queuing model is suggested in the current work. The queuing mechanism is developed based on the probability distribution in different range of communication. The queuing mechanism hence improves the network metrics such as overall network throughput, reduces the route delay, overhead and traffic blockage probability. The approach is generated over a routing scheme in adhoc network.

*Queuing Methodology Based Power Efficient Routing Protocol for Reliable Data Communications in Manets*

### 3 Routing Protocol

### 3.1 Dynamic Source Routing Protocol (DSR)

The dynamic Source Routing Protocol (DSR) is a simple and efficient routing protocol designed specifically for use in mobile nodes multi-hop wireless ad hoc networks. DSR allows network be completely self-organizing and self-configuring, without the need for an existing network infrastructure or administration. Network nodes cooperate packets to each other to allow communication over multiple "hops" between the nodes is not directly within the wireless range of transmission of the other. As the network nodes move or join or leave the network and that the conditions of wireless transmission, as sources of change in the interference, routing all is automatically determined and maintained by the DSR routing protocol. The DSR Protocol enables nodes to dynamically discover a source through several breaks of network route to any destination of the ad hoc network. Each packet of data sent and then carries in its complete header, ordered list of nodes which the packet must pass, for routing packets be trivialement free loop and avoiding the need for routing information updated in the intermediate nodes through which the packet is forwarded.

By including this source route in the header of each packet of data, other nodes to transfer or to hear any of these packages may also easily to cache thi s information in routing for future use. All aspects of the Protocol fully exploit to the application, which enables routing over DSR package scale automatically to only that he had to respond to changes in the routes currently in use. The Protocol comprises two mechanisms of discovery of the road and the maintenance of the road, working together to allow nodes to discover and maintain roads from the source to arbitrary destinations in ad hoc network.

### 3.2 Route-Discovery:

Consider a source node that does not have a route to the destination. When data packets must be sent to this destination, it initiates a Route request packet. This application for route is flooded throughout the system. Each node on receipt of a Route request packet retransmits the packet to its neighbors if it has not already passed or if the node is not the destination node, provided the time of the packet to live (TTL) counter has exceeded. Each request road is a sequence number generated by the source node and the path he crossed. A node, on receipt of a Route request packet, checks the sequence number of the packet before sending it. The package is passed only if it is not a request for duplicate route. The sequence number of the package is used to prevent loop formations and to avoid multiple transmissions of the same application Route by an intermediate node that it receives through multiple paths. Thus, all nodes except the destination send a request packet from the road during the route construction phase. A node destination, after the first packet to the road, responds to the source via the reverse path node that the motion of the road package had crossed. Nodes can learn on the neighboring routes traversed by packets of data if exploited in promiscuous mode ( the mode of operation in which a node can receive packets that are neither broadcast nor addressed to itself). The route cache is also used during the route construction phase. If an intermediate node receives a request for the road has a route to the destination in its route cache node, then it meets the source node by sending a response from the road with the information of all paths from the source to the destination node.

### 3.3 Route Maintenance

DSR each node is responsible for the confirming that the next hop in the route Source receives the packet. Each packet is transmitted only once by a node (hop-by-hop routing). If a packet can be received by a node, it is broadcast to some maximum number of times until the jump from the receipt of confirmation.

Only if the broadcast then causes a failure, a Route error message is sent to the initiator that can delete Source Route from its Route Cache. Thus, the initiator can check its Cache of route for another route to the target. If there is no route cache, a Route request packet is broadcasted.

*3.4 The problem Issues in Ad Hoc Management*

Mobile ad hoc networks (MANETS) consist of nodes which transmits data to other wireless form a connected network. These networks provide a rapid deployment and configuration auto functions and have applications In various environments such as battle fields, environmental monitoring and disaster recovery. Often the nodes in entangling limited energy supply. Thus, to increase the lifetime of network, a node must optimize its use of energy. In the communication system, the interface wireless between two nodes is the largest energy consumption [10]. The wireless interface consumes energy not only during an active communication, but also for passive listening, when it is inactive. Studies [4, 15] show that energy consumption while listening data is only slightly less it is receiving data. Thus, in the case of moderate traffic workload, idle time is the dominant factor in energy consumption.

    The other major factor in the Ad Hoc management is the bad behavior of node. Although a system of effective power management is applied to an ad hoc network to a low-power node may result in improper routing of packets that can extend up to the complete collapse of the network also. In mobile ad hoc networks, where nodes act as routers and terminals, nodes should cooperate to communicate. Cooperation at the level of the network layer takes place at the routing level , i.e. to find a path to a package and transfer, i.e. relay packets to other nodes. Misconduct means aberration of normal routing and transfer of behavior. Arise for several reasons.

    When a node is faulty , its erratic behavior may depart from the Protocol and thus produce a bad unintenTional behavior. Intentional bad behavior aims to provide a benefit for low-power node. An example for a Benefit obtained by the misconduct is power saved when a selfish node does not forward packets to other nodes. An advantage for a malicious node arises when the bad behavior allows mounting an attack. Without appropriate countermeasures, the effects of bad conduct demonstrated to significantly decrease network performance.

    Depending on the proportion of low-power nodes and their specific strategies, network throughput may be severely degraded, increased packet loss, nodes can be denied service and the network can be partitioned. These harmful effects of bad conduct can endanger the functioning of the network. The problem, we want to solve is, how make us an existing system continue to work despite the presence of bad conduct with a management regime effective power for ad hoc network of long life?

**4 Routing Management In Ad Hoc Network**

It is often necessary to manage a network ad hoc well due to the area in which these networks is used. For example, in a scenario disaster management software should provide a complete picture of the deployment of rescue teams, eventually covered of a map of the area containing information about the potential dangers, the density of the victim, etc. In this scenario, we can also imagine data from sensors in UN populated (on the ground, as in the Earth earthquake struck areas or airborne, as in the case of nuclear disaster) being relied on the management station that provide a complete picture of the situation.

*4.1Issues in Routing Management*

The absence of a central infrastructure means that an ad hoc network does not have a fixed topology partner. Indeed, an important task of an ad hoc network consisting of geographically dispersed nodes is to determine an appropriate topology in which high level routing protocols are implemented. In this section, we consider the management of the topology, the problem of the determination of a topology appropriate in an ad hoc network.

*4.2Connectivity and energy-efficiency*

The most basic requirement of the topology is perhaps that it is connected. We wish to provide the connectivity and efficiency with a "simple" "easy" to keep topology. "While there is no single way to formalize the "simplicity" and "maintenance", some objective measures that influence these subjective goals is the size of the topology in terms of the MARI level of nodes. What distinguishes the topology management problem in ad hoc mobile of traditional network design parameter is that we must determine the topology in a completely distributed environment. Thus every node in an ad hoc network must make decisions locally based on information obtained from the neighbors.

*4.3 Throughput*

In addition to connectivity and energy efficiency, we would like to have a topology high capacity or throughput . i.e., it must be possible to route "about as much traffic" in the topology as any other topology satisfying the desired constraints. The number of feasible particular interference, however, depends on the relative position of the ad hoc network nodes and their transmission rays. This brings us to the following problem in network design: given a collection of node s in ad hoc network, design a connected topology that minimizes the number of interference. It seems unlikely that the previous optimization problem can be solved efficiently by a local algorithm; however, an algorithm for the problem may be of theoretical interest.

*4.4 Robustness to mobility*

An additional challenge in the design of algorithms of distributed management topology is to ensure certain solidity to the mobility of the nodes. A measure of the strength of the topology is given but the number maximum of nodes that need to modify their information of topology of a movement of a node. The number, which may be mentioned as the adaptability of the algorithm of topology management depends on the size of the area of transmission of the mobile node u, and the location of the nodes, other than the maintenance of the topology, mobility also results in changes in the routing paths.
The Topology Management in Ad hoc Wireless Networks id deciding for every node:

- Which node to turn on.
- When they turn on.
- At what transmit MARI.

So that network connectivity is maintained under the conditions of mobility.

## 5. QUEUING SYSTEMS

A queuing system consists of one or more servers that provide service of some sort to arriving customers. Customers who arrive to find all servers busy generally join one or more queues (lines) in front of the servers, hence the name queuing systems. There are several everyday examples that can be described as queuing systems [7], such as bank-teller service, computer systems, manufacturing systems, maintenance systems, communications systems and so on.
   Components of a Queuing System: A queuing system is characterized by three components:
Arrival process - Service mechanism - Queue discipline.

a)Arrival Process
   Arrivals may originate from one or several sources referred to as the calling population . The calling Population can be limited or 'unlimited'. An example of a limited calling population may be that of a fixed number of machines that fail randomly. The arrival process consists of describing how customers arrive to the system. If $A_i$ is the inter-arrival time between the arrivals of the (i-1)th and *i*th customers, we shall denote the mean (or expected) inter-arrival time by $E(A)$ and call it $(\lambda)$; $= 1/E(A)$ the arrival frequency.
b)Service Mechanism
   The service mechanism of a queuing system is specified by the number of servers (denoted by s), each server having its own queue or a common queue and the probability distribution of customer's service time.

Let Si be the service time of the ith customer, we shall denote the mean service time of a customer by E(S) and μ = 1/(E(S)) the service rate of a server.

c) Queue Discipline

Discipline of a queuing system means the rule that a server uses to choose the next customer from the queue (if any) when the server completes the service of the current customer. Commonly used queue disciplines are:

FIFO - Customers are served on a first-in first-out basis. LIFO - Customers are served in a last-in first-out manner. Priority - Customers are served in order of their importance on the basis of their service requirements.

d) Measures of Performance for Queuing Systems:

There are many possible measures of performance for queuing systems. Only some of these will be discussed here.

Let, Di be the delay in queue of the ith customer Wi be the waiting time in the system of the ith customer = Di + Si Q(t) be the number of customers in queue at time t L(t) be the number of customers in the system at time t = Q(t) + No. of customers being served at t.

Then the measures,

$$d = Lim_{x \to \infty} \frac{\sum_{i=1}^{i=n} D_i}{n} \quad \text{and}$$

$$w = Lim_{x \to \infty} \frac{\sum_{i=1}^{i=n} W_i}{n} \tag{1}$$

(if they exist) are called the steady state average delay and the steady state average waiting time in the system. Similarly, the measures,

$$Q = Lim_{T \to \infty} \frac{1}{T} \int_0^T Q(t).dt \quad \text{and}$$

$$L = Lim_{T \to \infty} \frac{1}{T} \int_0^T L(t).dt \tag{2}$$

(if they exist) are called the steady state time average number in queue and the steady state time average number in the system. Among the most general and useful results of a queuing system are the conservation equations:

$$Q = (\lambda)d \quad \text{and} \quad L = (\lambda)w \tag{3}$$

These equations hold for every queuing system for which d and w exist. Another equation of considerable practical value is given by,

$$w = d + E(S) \tag{4}$$

Other performance measures are:

the probability that any delay will occur - the probability that the total delay will be greater than some pre-determined value - that probability that all service facilities will be idle - the expected idle time of the total facility - the probability of turn-a ways, due to insufficient waiting accommodation.

e) Notation for Queues.

Since all queues are characterized by arrival, service and queue and its discipline, the queue system is usually described in shorten form by using these characteristics. The general notation is:

$$[A/B/s]:\{d/e/f\}$$

*Queuing Methodology Based Power Efficient Routing Protocol for Reliable Data Communications in Manets*

Where,

A = Probability distribution of the arrivals
B = Probability distribution of the departures
s = Number of servers (channels)
d = The capacity of the queue(s)
e = The size of the calling population
f = Queue ranking rule (Ordering of the queue)

There are some special notation that has been developed for various probability distributions describing the arrivals and departures. Some examples are,

M = Arrival or departure distribution that is a Poisson process
E = Erlang distribution
G = General distribution
GI = General independent distribution

Thus for example, the $[M/M/1]$: {infinity/infinity/FCFS} system is one where the arrivals and departures are a Poisson distribution with a single server, infinite queue length, calling population infinite and the queue discipline is FCFS. This is the simplest queue system that can be studied mathematically. This queue system is also simply referred to as the *M/M/1* queue.

## 6. RESULT ANALYSIS

In order to achieve the objective of this work a network is created with 30 nodes (while simulation nodes can be varied) and routing protocol l is applied to the network and the performance metrics are evaluated as shown below. The environment is created in such a way that the nodes are mobile and while mobility the parameters should be maintained effectively. Hence we adopted queuing methodology additional to proposed routing protocol to get the reliability in communication. As queuing methodology is a mathematical random and poisson process it will provide discipline in communication. It will have control over service rate and arrival rate of data packets due to its server models and maintenance of queue without any congestion. The figure below illustrates the nodes in topology formation.

Created Network : 30 Nodes

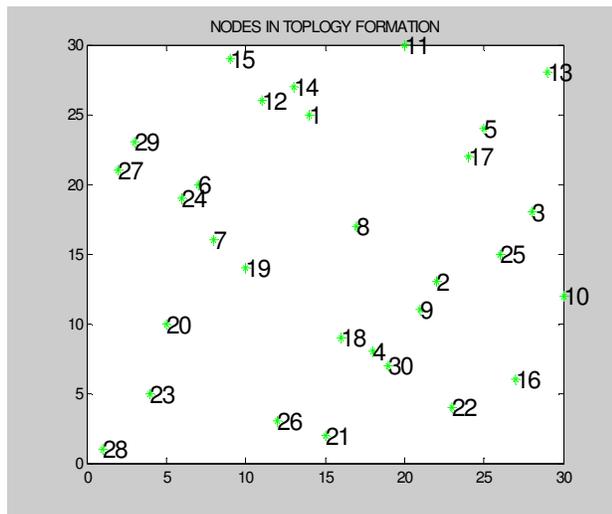

**Figure 1.** Nodes in topology formation

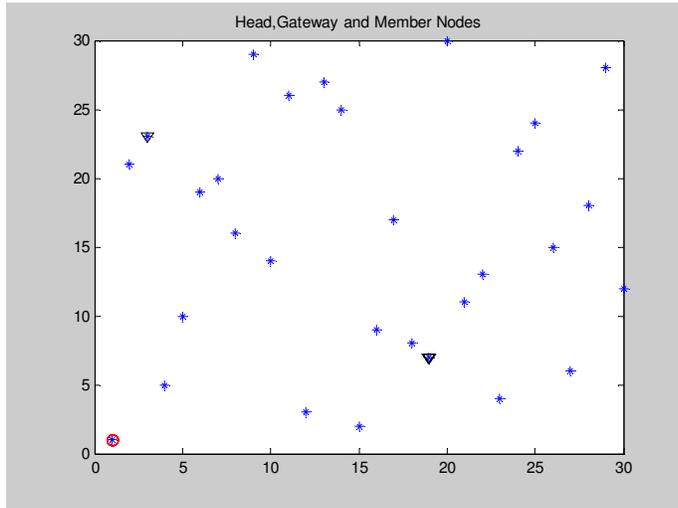

**Figure 2.** Declared head, member and gateway

The above figure consists of nodes in the environment which will be changing adaptively. We have to declare head, member and gateway node to start communication. Those are named and situated. The communication happens through the specified nodes according to the specifications of the standards.
For a node density = 30 nodes (Average Node Density Network)
Observation communication period = 5

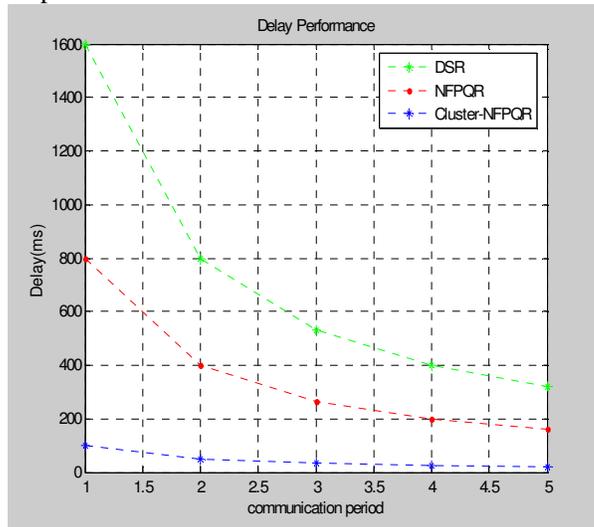

**Figure 3.** Delay performance

In this work we have done a review of routing protocols for the performance estimation. In the process it is found that the delay performance for cluster NFPQR is less as the communication period increases which implies that after applying queuing methodology the congestion is decreased and the service rate as well arrival rates are maintained properly.

The above plots indicate that the enhanced routing protocol has got good throughput, efficient power consumption and increased network lifetime. The overhead per node also decreased due to the enhancement in routing protocol and the application of queuing methodology.

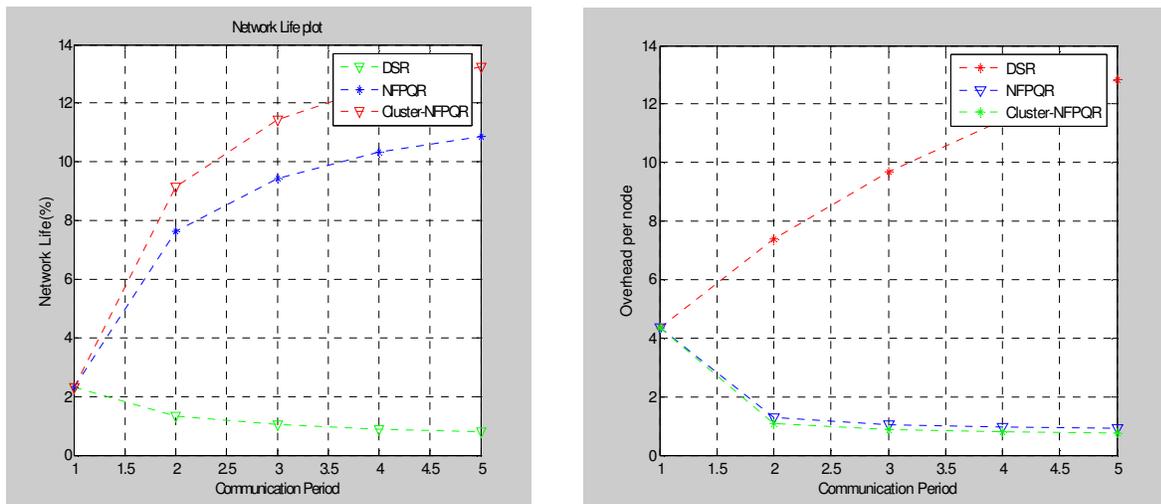

**Figure 4.** Network life and overhead per node

## 7. Conclusions

In this paper, we proposed an efficient secure routing protocol for mobile ad hoc networks that guarantees the discovery of correct connectivity information over an unknown network, in the presence of malicious node. The protocol introduces a set of features, such as the requirement that the query verifiably arrives at the destination, the explicit binding of network and routing layer functionality, the consequent verifiable return of the query response over the reverse of the query propagation route, the acceptance of route error messages only when generated by nodes on the actual route, the query/reply identification by a dual identifier, the replay protection of the source and destination nodes and the regulation of the query propagation. The resultant protocol is capable of operating without the existence of an on-line certification authority or the complete knowledge of keys of all network nodes. Its sole requirement is that any two nodes that wish to communicate securely can simply establish a priori a shared secret, to be used by their routing protocol modules. By introducing queuing methodology to the proposed routing protocol the reliability of the selected protocol increased and throughput of the system is also maximized.